\documentclass[10pt,journal,compsoc]{IEEEtran}
\pdfoutput=1

\ifCLASSOPTIONcompsoc
  \usepackage[nocompress]{cite}
\else
  \usepackage{cite}
\fi

\usepackage{cite}
\usepackage{amsmath,amssymb,amsfonts}
\usepackage{algorithmic}
\usepackage{graphicx}
\usepackage{textcomp}
\usepackage{adjustbox}
\usepackage{xcolor}

\definecolor{black}{rgb}		{0.0, 0.0, 0.0}
\definecolor{white}{rgb}		{1.0, 1.0, 1.0}
\definecolor{yellow}{rgb}		{1.0, 1.0, 0.8}
\definecolor{red}{rgb}			{0.6, 0.0, 0.2}
\definecolor{blue}{rgb}		{0.0, 0.2, 0.5}
\definecolor{green}{rgb}		{0.6, 0.8, 0.8}
\definecolor{dark_green}{RGB} {0, 140, 0}
\definecolor{gold}{rgb}		{0.6, 0.4, 0.1}
\definecolor{grey}{RGB}{0,0,0}
\definecolor{Gray}{gray}{0.8}
\definecolor{MediumGray}{gray}{0.9}
\definecolor{LightGray}{gray}{0.98}
\definecolor{LightCyan}{rgb}{0.88,1,1}
\definecolor{purple}{RGB}{128,0,128}
\definecolor{sl_blue}{RGB}{47, 60, 105}
\definecolor{orange}{RGB}{255,165,0}
\definecolor{Gray}{gray}{0.85}

\usepackage{subcaption}
\usepackage{caption}
\usepackage{url,times,booktabs, tabularx}
\usepackage[colorlinks=false,pdfborder={0 0 0}]{hyperref}
\usepackage{units}
\usepackage{multirow,array}
\usepackage{balance}
\usepackage{float}
\usepackage{soul}
\usepackage{color}
\usepackage{cleveref}
\usepackage{wrapfig}
\usepackage{booktabs}
\usepackage[activate]{microtype}
\usepackage{array}
\newcolumntype{P}[1]{>{\centering\arraybackslash}p{#1}}
\newcommand{\eg}{e.\,g.,\ }
\newcommand{\ie}{i.\,e.,\ }

\newcommand{\cf}{{cf.\,}}

\usepackage{soul}
\usepackage{xcolor}

\usepackage{dirtytalk}

\begin{document}
%
\title{Robust Federated Learning Against Adversarial Attacks for Speech Emotion Recognition}
%
%
%
%

\author{Yi~Chang,~\IEEEmembership{student member,~IEEE}, Sofiane~Laridi, Zhao~Ren,~\IEEEmembership{member,~IEEE}, Gregory~Palmer, Björn~W.~Schuller,~\IEEEmembership{Fellow,~IEEE}, Marco~Fisichella
\IEEEcompsocitemizethanks{
\IEEEcompsocthanksitem Y.\ Chang and B.\ W.\ Schuller are with GLAM -- the Group on Language, Audio, \& Music, Imperial College London, SW7 2AZ London, UK. Email: y.chang20@imperial.ac.uk, schuller@ieee.org
\IEEEcompsocthanksitem S.\ Laridi, Z.\ Ren, G.\ Palmer, and M.\ Fisichella are with the L3S Research Center, Leibniz University Hannover, 30167 Hannover, Germany. Email: \{laridi, zren, gpalmer, mfisichella\}@l3s.de
\IEEEcompsocthanksitem B.\ W.\ Schuller is also with the Chair of Embedded Intelligence for Health Care and Wellbeing, University of Augsburg, 86159 Augsburg, Germany.
}
\thanks{All authors provided critical feedback and helped shape the research. 
Y.\ Chang and S.\ Laridi contributed equally to the development and evaluation of the proposed methodology. Z.\ Ren, G.\ Palmer, and M.\ Fisichella equally conceived the original idea and experimental design. Z.\ Ren, G.\ Palmer, M.\ Fisichella, and B. W. Schuller reviewed the methodology. M.\ Fisichella coordinated and supervised the project.}
\thanks{Corresponding author: Z.\ Ren.}}

%
%

\markboth{XX,~Vol.~XX, No.~XX, XX~2022}%
{Shell \MakeLowercase{\textit{et al.}}: Bare Demo of IEEEtran.cls for Computer Society Journals}
%



\IEEEtitleabstractindextext{
\begin{abstract}
Due to the development of machine learning
and speech processing, 
speech emotion recognition has been a popular research topic in recent years. However, the speech data cannot be protected when it is uploaded and processed on servers in the internet-of-things applications of speech emotion recognition. Furthermore, deep neural networks have proven to be vulnerable to human-indistinguishable adversarial perturbations. The adversarial attacks generated from the perturbations may result in deep neural networks wrongly predicting the emotional states. We propose a novel federated adversarial learning framework for protecting both data and deep neural networks. The proposed framework consists of i) federated learning for data privacy, and ii) adversarial training at the training stage and randomisation at the testing stage for model robustness. The experiments show that our proposed framework can effectively protect the speech data locally and improve the model robustness against a series of adversarial attacks.

\end{abstract}

}

\maketitle

\IEEEdisplaynontitleabstractindextext

%
\IEEEpeerreviewmaketitle

\IEEEraisesectionheading{\section{Introduction}\label{sec:introduction}}

\IEEEPARstart{V}{oice} operated smart edge devices -- \eg Amazon Alexa and Google Nest 
-- are becoming increasingly prevalent in our daily lives. 
These devices have benefited from recent advancements within the field of
Machine Learning (ML), where deep neural networks
represent the current state-of-the-art approach for analysing 
speech and recognising emotional states~\cite{khalil2019speech}. 
In this paper, we address two topical challenges within this area:
\emph{data privacy} and 
\emph{robustness
towards adversarial attacks}.

Despite recent advances, the performance of ML models remains upper-bounded 
by the quality and quantity of the samples used for training~\cite{jain2020overview}. 
This limitation has implications for the field of Speech Emotion Recognition (SER).
The data used in this area often contains private / sensitive information, especially when emotion recognition is applied in mental health applications, \eg detection of depression~\cite{gao2018machine}. 
Therefore, while smart edge devices may record an abundance of data, 
there are concerns regarding the extent to which 
these devices may leak private information~\cite{9546911}. 
These concerns -- along with strict legal and ethical requirements 
designed to protect user privacy -- 
can prohibit the pooling of user data for \emph{centralised} ML model training, 
a practice that often results in a better model. 
Thus, due to the lack of data from different sources, 
it can become a challenge to build ML models that are effective
for SER. 
Here, Federated Learning (FL) has recently emerged as an alternative to centralised 
learning, allowing data holders to collaboratively train a \emph{global} model without physically sharing 
their data. 
Instead, participants -- in our case, the users of the smart edge devices -- train a 
copy of a model with local data and iteratively share the resulting parameter 
updates, often via a centralised entity~\cite{li2020federated}. 
Therefore, thanks to FL, a \emph{global} model can be obtained without the users' data 
leaving their respective devices. 

While FL takes steps towards mitigating privacy concerns when training
models using real-world data, questions remain regarding model robustness.
In recent years, there has been an abundance of literature
demonstrating the vulnerability of deep neural networks towards adversarial
attacks~\cite{narodytska2017simple}. 
Adversaries are capable of learning to perturb a small set of pixels
within a given sample, barely perceptible to humans, but capable
of causing a misclassification.
Such attacks are increasingly being used to fool models deployed on smart devices,
often with malicious intent~\cite{9130128}. 
This creates the need for models that are robust towards adversarial attacks, 
as well as principled methodologies for measuring robustness. 

In this paper, we take steps towards addressing the challenges outlined above.
To the best of the authors' knowledge, we propose the first pipeline to protect federated learning with a comprehensive approach in the domain of SER. Our contributions can be summarised as follows:

i.) A federated learning framework for SER is constructed to protect each speaker's data privacy. 

ii.) To protect SER models trained with federated learning against a range of adversarial attacks, we propose and compare two single-stage federated defence modelling strategies: adversarial training at the training stage and randomisation at the testing stage. Especially, our experiments demonstrate that randomisation on log Mel spectrograms extracted from speech signals is able to protect the SER models against adversarial attacks.

iii.) Finally, we show that our proposed two-stage defence modelling approach, \ie the combination of adversarial training and randomisation, effectively improves the model robustness over vanilla federated learning (\ie nature) and federated learning with a single stage of defence.

\section{Background}
 \label{sec:background}

This chapter will go through the basic principles of SER and FL. We also provide a recap on adversarial attacks and their types, followed by two techniques to defend against these attacks, namely: i)~adversarial training for the single-step adversarial white-box attacks; and ii)~randomisation to defend against iterative adversarial white-box attacks.

\subsection{Speech Emotion Recognition}

Human-Computer Interaction (HCI) is an essential part of Artificial Intelligence (AI) research.
Real-life HCI applications have been facilitated by research on automatic emotion recognition~\cite{sharma2021survey},  
thus improving quality of service and quality of life~\cite{weninger2015emotion}. 
The speech signal is one of the key ways for humans to communicate, since it contains a substantial amount of paralinguistic information (\ie emotion states, attitudes, etc.)~\cite{jovicic2004serbian}. In past decades, SER has been a particularly useful research topic for HCI to recognise emotion from speech signals with the methods of computational paralinguistics and machine learning~\cite{uar1}. In SER tasks, speech signals are often annotated with continuous values (\ie \emph{arousal}, \emph{valence} and \emph{dominance}~\cite{busso2008iemocap}) and/or discrete labels (\eg the `Big Six' proposed by Ekman~\cite{ekman1984expression}: \emph{anger}, \emph{disgust}, \emph{fear}, \emph{happiness}, \emph{sadness} and \emph{surprise}). 
The continuous and discrete labels can be converted to each other using a circumplex model~\cite{russell1980circumplex}. 

In the past years, a set of acoustic features have proven to be effective for distinguishing emotions from speech, including low-level descriptors (\eg energy features and spectral features) and statistical functionals~\cite{eyben2010opensmile,eyben2015geneva}. With acoustic features, classic machine learning approaches (\eg support vector machines) have been successfully employed to analyse emotional states from speech signals~\cite{schuller2020interspeech}. In contrast to classic machine learning, deep learning 
mostly 
deals with either raw speech signals or time-frequency representations, and often shows better performance than classic machine learning in recent advances~\cite{gerczuk2021emonet,tzirakis2018end}. Compared to 1D raw speech signals, 2D time-frequency representations have become more popular~\cite{zhang2021learning}. Typical time-frequency representations include spectrograms, Mel spectrograms, log Mel spectrograms, and Mel Frequency Cepstral Coefficients (MFCCs)~\cite{badshah2017speech,ren2020generating,wang2020speech}. In this work, we will use log Mel spectrograms as the input of Convolutional Neural Networks (CNNs) due to the Mel scale's linearity in the human ear's auditory properties~\cite{xue2020speaker} and log Mel spectrograms' good performance as the input of CNNs in acoustic tasks~\cite{ren2020caanet}.


\subsection{Federated Learning}
\label{sec:fl}

FL has become a popular solution for applications
that can benefit from a decentralized, 
collaborative and privacy preserving learning process.
Application domains that can benefit from an FL approach include: 
intelligent industrial production~\cite{ge2021failure}, 
healthcare \cite{xu2021federated} and, the focus of our current study, SER~\cite{9130128}.
FL consists of a number of client machines 
hosting a decentralized training dataset,
where each sub-dataset 
resides and remains on the respective client machine
throughout the training process~\cite{bonawitz2019towards}.
Instead of sharing data, the clients share the parameters 
of an ML model, typically via a centralized server
hosting a \emph{global model}, or via a decentralized 
peer-to-peer topology~\cite{li2020federated}. 
This global model can be downloaded by each client
to perform \emph{local} updates. 
After a specified number of epochs, each client sends 
their optimized model parameters back to the server,
where they are aggregated -- for instance using federated averaging~\cite{konevcny2016federated}. 
The aggregated parameters are subsequently made available for
download, and the next training \emph{round} can begin. 
%

\subsection{White-box Adversarial Attacks}

The digital transformation in mobile networks and computing has recently led to a dramatic increase in the number of Internet users, connections, and Internet-of-Things (IoT) devices, as well as network capabilities and application requirements. 
With the digital proliferation, 
increasingly sophisticated attack strategies
are  
emerging to take the offensive against Deep Neural Networks (DNNs). 
A thorough investigation has been conducted on the vulnerability of DNNs to various malicious attacks capable of creating adversarial examples with the goal of breaking the prediction of DNNs~\cite{10.1007/978-3-030-62460-6_51}. 
There are two major categories of these attacks: 
1) \emph{white-box attacks} that consider the available model architecture; 
2) \emph{black-box attacks} without any knowledge of the model architecture.
Regardless of the category, the goal of most attacks is to perturb the inputs of DNNs with so-called adversarial examples in such a way that the new examples are indistinguishable to humans. The new false examples can mislead the prediction model and produce incorrect results. Compared with black-box attacks, white-box attacks tend to generate more imperceptible and thus stronger perturbations. Therefore, we focus on the more challenging white-box attack~\cite{Ren2020EnhancingTO}.    

In this paper, we investigate the extent to which we can 
design robust defence schemes for models obtained via FL against the following adversarial white-box attacks: 

\noindent{\emph{The Fast Gradient Sign Method} (FGSM)} generates an adversarial sample $x_{adv}$ from an individual input sample $x \in \mathcal{X}$ (where $\mathcal{X}$ represents a set of samples) in a single step, using the sign
of the gradient $\nabla$ of a loss function $\mathcal{L}$ of the network~\cite{advexample2015}:
\begin{equation} \label{eq:FGSM}
x_{adv} = x + \epsilon\ \mbox{sign}(\nabla_x\mathcal{L}(\theta,x,y)),
\end{equation}
where $\epsilon$ is a positive constant that determines the scale of the perturbation , and $\theta$ denotes the model parameters.

\

\noindent{\emph{Projected Gradient Attack} (PGD)~\cite{madry2018towards,liu2019adversarial}} is a more powerful multi-step variant of FGSM:

\begin{equation} \label{eq:PGD}
x^{i+1} = \mbox{clip}(x^i + \alpha\ \mbox{sign}(\nabla_{x^i}\mathcal{L}(\theta,x^i,y)),
\end{equation}
where $i$ is the iteration step, and $\alpha$ is a constant value that affects the perturbation's scale. The $\mbox{clip}(\cdot)$ function aims to clip the perturbations into a small interval, \ie $|\alpha\ \mbox{sign}(\nabla_{x^i}\mathcal{L}(\theta,x^i,y))|\leq \eta$, where $\eta$ is a positive constant. Compared to FGSM, PGD yields a number of possible adversarial samples in multiple iteration steps, but it has a higher computational complexity~\cite{liu2019adversarial}.

\


\noindent{\emph{DeepFool}}~\cite{moosavi2016deepfool} is a typical iterative attack, and aims to find the minimal perturbation $r$ that results in a change in output 
of a classifier $\mathcal{M}(\cdot)$ when applied to a sample $x$.
The assumption is made that $\mathcal{M}(\cdot)$ is an affine binary classification function.
DeepFool estimates the perpendicular distance and direction from an input $x$ to the decision boundary $\mathcal{M}(x)=0$. 
A constant $(1 + \zeta)$ is used to multiply $r$ to generate $x_{adv}$, ensuring that the $\mathcal{M}(x_{adv})$ crosses the decision boundary:  
\begin{equation} \label{eq:DeepFool}
\delta = r \times (1 + \zeta),   \quad \quad   x_{adv} = x + \delta.
\end{equation}
Correspondingly, DeepFool in binary classification can be extended to multi-class classification according to the one-vs-all classification scheme~\cite{moosavi2016deepfool}.

\subsection{Adversarial Training}
\label{sec:adv_training}

Adversarial examples reveal the inherent vulnerability of neural networks due to their linear nature~\cite{advexample2015}. In order to fight against the adversarial examples, adversarial training was proposed and initially utilised to train the neural networks with a mixture of adversarial examples and original clean signals to improve the robustness of neural networks~\cite{advtrain2014, KurakinGB17}. Ensemble adversarial training was proposed in~\cite{tramer2020ensemble} to further augment training data with perturbations generated from other pre-trained models. 
Max-Margin Adversarial (MMA) was proposed in~\cite{ding2020mma} to directly maximise the margins to the decision boundary and minimise the adversarial loss on the decision boundary at the \say{shortest successful perturbation}. Moreover, adversarial robustness can also be achieved by including additional unlabelled data and thereby reducing the sample complexity gap between adversarial and clean samples~\cite{3455291, 3455381}.

Besides the adversarial training, denoising techniques can also be applied to increase the model's robustness toward adversarial attacks, such as a  high-level representation guided denoiser~\cite{Liao_2018_CVPR} and feature denoising~\cite{Xie_2019_CVPR}. However, these denoising techniques either under-perform because the defence can be circumvented~\cite{DBLP:conf/icml/AthalyeC018} or include extra parameters to train.  
\subsection{Randomisation} \label{sec:randomisation}

Randomisation was introduced in~\cite{xie2018mitigating} as a defence method at the inference time against the adversarial attacks. It contains two layers: random resizing and random padding as indicated in Equation~(\ref{eq:randomisation_org}).
The random resizing layer randomly enlarges the input images with the dimension of $(W, H, C)$ ($W$: Width, $H$: Height, $C$: Channel number) to the dimension of $(W', H', C)$, which is further padded into the dimension of $(W'', H'', C)$. 
Through fine-tuning the hyper-parameters during the resizing and padding procedures, randomisation can work effectively to mitigate the adversarial effects.
\begin{equation} \label{eq:randomisation_org}
(W, H, C) \xrightarrow[\mbox{resizing}]{\mbox{random}} (W', H', C) \xrightarrow[\mbox{padding}]{\mbox{random}} (W'', H'', C).
\end{equation}
%
%
Whereas adversarial training can effectively improve the models' robustness against single-step attacks (\eg FGSM)~\cite{tramer2020ensemble, KurakinGB17}, image transformations (\eg resizing, padding) have been found to mitigate the effects of iterative adversarial attacks (\eg PGD, DeepFool) due to their weak generalisation ability~\cite{xie2018mitigating}.
In this work, we also combine adversarial training with randomisation to better mitigate the effects of single-step as well as iterative adversarial attacks.


\section{Related Work} \label{sec:related_work}

While there exist a few research studies on federated learning for SER, as well as adversarial attack on SER in centralised settings,
to the best of the authors' knowledge, our work is the first exploration of improving federated learning's capability for privacy protection against adversarial attacks in this area. 
%
%
The potential benefits of FL for SER's typically sensitive data -- and ability to deliver promising results compared with SOTA approaches -- were highlighted in a recent study by~\cite{latif2020federated}.
A number of real world applications have been identified for Federated-SER, including: monitoring the general mood of elderly in care homes~\cite{das2019privacy}; the healthcare industry in general, a vulnerable sector that often falls victim to cyber-attacks and data breaches, where data is typically highly sensitive and distributed in nature~\cite{elayan2021sustainability}; multi-modal (\ie face video and speech) emotion recognition systems to improve work culture and the environment in
post-pandemic times~\cite{chhikara2020federated}; and depression treatment robots that do not need to transfer users' videos and conversation data to the server~\cite{liu2021federated}.

We note that there have also been recent works focused on \emph{privacy attacks} on Federated-SER, where the goal is to retrieve / recover sensitive information from trained models, typically as a result of the models over-fitting on the training data\cite{tomashenko2021privacy,das2019privacy}.
A further concern are model \emph{poisoning attacks}, where FL participants deliberately attempt to sabotage the model so that
it misclassifies specific samples, potentially those of relevance to competitors~\cite{zhang2019poisoning}.
Model poisoning attacks differ from the white-box attacks that are the focus of our current work, since for white-box attacks, the goal is to perturb samples to cause a misclassification at \emph{inference time}.

A comprehensive summary of FL challenges -- including communication costs, resource allocation, privacy and security -- can be found in a survey~\cite{9060868}.
%
%
Meanwhile, for SER in centralised settings, there exists a number of works on adversarial attacks and corresponding defence schemes.
An end-to-end scheme was proposed to generate adversarial emotional speech data in~\cite{gong2017crafting}. 
In~\cite{latif2018adversarial}, the first black-box adversarial attacks were used to deceive SER systems, and adversarial training and Generative Adversarial Networks (GANs) 
were explored to enhance the SER model's robustness. 
Furthermore, a similarity-based adversarial training was proposed to protect SER models against white-box adversarial attacks in our prior study~\cite{ren2020generating}. 
As improving the transferability of adversarial attacks can facilitate the investigation of improving SER models' robustness, the transferability of black-box adversarial attacks was enhanced by lifelong learning in another study~\cite{Ren2020EnhancingTO}. 

There have also been numerous studies on defensive methods, \eg adversarial training and ensemble diversity~\cite{pang2019improving}. 
However, most of them are not suitable for distributed devices or for distributed machine learning. 
The main reason is that the devices are scattered and may be exposed to different attacks simultaneously. 
%
%
%
The first attempt to adopt the concept of FL to provide a defence framework against various white-box attacks on distributed networks was studied in~\cite{9130128}. 
The authors present the $FDA^3$ approach, which is able to pool the knowledge of defending against attacks from different sources.
However, the approach assumes an \emph{attack monitor} capable of determining which type of white-box attack is being applied.
In contrast, our current work focuses on mitigating the effects of adversarial attacks on Federated-SER at inference time through randomisation, an approach that works well on image data~\cite{xie2018mitigating}. 


\section{Problem Description} \label{sec:problem_desc}

In this section, we formally define our problem setting, 
introducing the terminology and notations that will be used
in the remainder of the paper. 
In particular, we formally define the FL scenario that we are 
studying as well as the application of white-box 
attacks. 
In Table~\ref{tab:notations}, we provide a summary of the notations used.

\begin{table*}[]
    \centering
    \caption{The descriptions of the notations used in federated learning.}
    \begin{tabular}{c|c}
    \toprule
         \textbf{Notations} & \textbf{Description}   \\
        
         $(\mathcal{X}, \mathcal{Y})$ & A set of samples $\mathcal{X}$ with corresponding labels $\mathcal{Y}$. \\
        \hline
        $(x,y )$ &
        We define an individual sample-label pair as $(x, y)$ where $x \in \mathcal{X}$ and $y \in \mathcal{Y}$ respectively. \\        
        \hline
         $I$ & Sample IDs. \\
        \hline                
         $\mathcal{F}$ & A set of data owners (federated learning \emph{clients}). \\
        \hline
         $\mathcal{D}_i$ & Dataset, or in the context of federated learning a set of datasets, where each $\mathcal{D}_i$ belongs to a data owner $\mathcal{F}_i$. \\
        \hline
         $\mathcal{M}$ & A machine learning model such as a deep neural network. \\
        \hline        
         $\theta$, $\vartheta$ & Parameters (weights) of a machine learning model. \\
        \hline   
         $\mathcal{V}$ & Performance measure for a machine learning model $\mathcal{M}$, for instance accuracy. \\
    \bottomrule
    \end{tabular}
    \label{tab:notations}
\end{table*}

\textbf{Federated Learning Scenario:} We assume a FL setting with $N$ data owners 
$\{\mathcal{F}_1, ... , \mathcal{F}_N\}$
and  their respective sub-datasets 
$\{\mathcal{D}_1, ... , \mathcal{D}_N\}$.
Each $\mathcal{D}_i$ contains a set of samples
$\mathcal{X}_i$ and corresponding labels $\mathcal{Y}_i$ for clients  $i\in[1,N]$~\cite{yang2019federated}.
A conventional approach towards ML model training would be to pool these sub-datasets
$\mathcal{D}_1 \cup	\mathcal{D}_1 ... \cup \mathcal{D}_N$ 
and train a model $\mathcal{M}_{\vartheta}$.
However, in FL, the data owners $\mathcal{F}_i$ do not wish to share their data.
Instead, each participant $\mathcal{F}_i$ obtains and updates a 
local copy of a model $\mathcal{M}_{\theta}$.
Here, $\vartheta$ and $\theta$ represent ML model parameters 
(weights) for a conventional (centralised) and federated models respectively, which are optimised during training.
Each participant $\mathcal{F}_i$ uses its respective sub-dataset $\mathcal{D}_i$, 
to update the local copy of $\mathcal{M}_{\theta,t}$'s parameters (weights) $\theta_{i,t}$, 
where $t$ represents
the current FL \emph{round}. 
The local model $\mathcal{M}_{\theta_{i,t}}$ is updated for $n$ epochs.
Then, the updated model parameters $\theta_{i, t}$ are sent to the FL server and aggregated. This gives us a new aggregated model $\mathcal{M}_{\theta_{t+1}}$ that is sent back to the FL participants , and the process repeats.

%

\textbf{White-box attacks on federated learning models:} We shall assume that our 
FL model $\mathcal{M}_{\theta}$ is vulnerable towards white-box
attacks, \eg someone has obtained a copy of the trained model~$\mathcal{M}_{\theta}$,
and can generate adversarial samples $x_{adv,i} \in \mathcal{X}_{adv}$ that would cause a 
misclassification for $\mathcal{M}_{\theta}(x_{adv,i})$, where $i$ represents the ID of the original (unperturbed) sample $x_i$.

\textbf{Performance Measure:} The goal in FL is to obtain a model $\mathcal{M}_{\theta}$
that achieves a comparable performance $\mathcal{V}_{\theta}$  (\eg with respect to accuracy) to what can be achieved using conventional centralised training.
The key distinction is being that during FL training, the data $\mathcal{D}_i$ physically remains with participant $\mathcal{F}_i$, thereby preserving privacy during 
the model training process~\footnote{While overall there 
are fewer privacy-specific threats associated with FL, the models themselves have the same vulnerabilities towards 
inference-based attacks as 
models obtained via conventional 
training~\cite{shokri2017membership}. 
This topic however, is outside of the scope of our current work. 
However, for readers who 
are interested in this topic, we recommend a recent survey \cite{mothukuri2021survey}.}.
Formally, the  objective of FL is to obtain a minimal $\delta$ 
accuracy loss: 
$| \mathcal{V}_{\theta} - \mathcal{V}_{\vartheta}| < \delta$~\cite{yang2019federated}.
In our problem setting, we also have the challenge of 
adversarial white-box attacks. 
Therefore, with a slight abuse of terminology, we also 
want to minimise:
$| \mathcal{V}_{\theta,\mathcal{X}_{adv}} - \mathcal{V}_{\vartheta,\mathcal{X}}| < \delta$, where $\mathcal{V}_{\theta,\mathcal{X}_{adv}}$ is the performance
of the model $\mathcal{M}_\theta$ when given a set of perturbed samples $\mathcal{X}_{adv}$ obtained via a white-box attack.

\section{The Proposed Methodology} \label{sec:method}

 \begin{figure*}
    \centering
    \includegraphics[width=.98\textwidth]{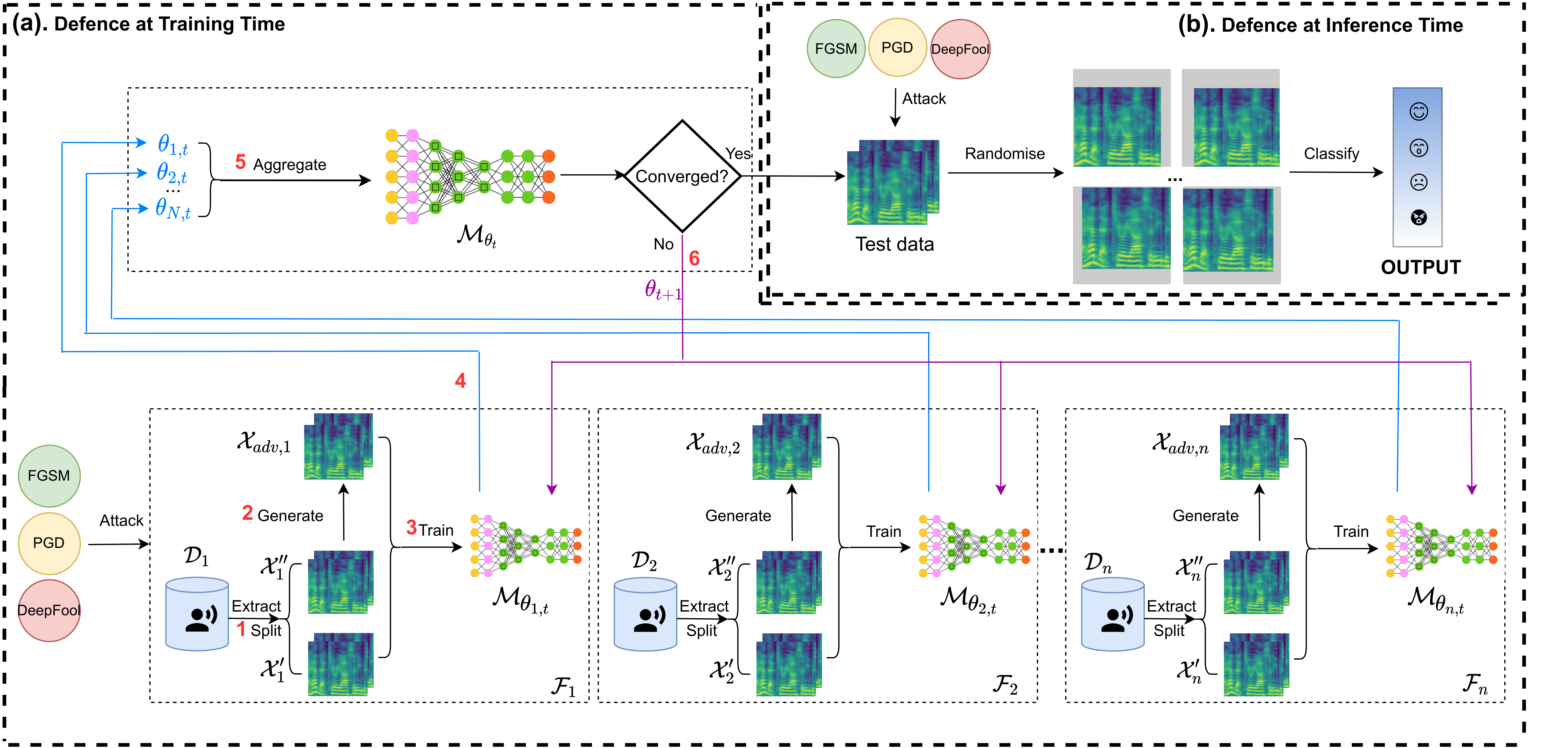}
    \caption{Overview of the proposed framework. Please refer to Table~\ref{tab:notations} for notations.
    1.) Split data $\mathcal{X}_i$ to $\mathcal{X}'_i$ \& $\mathcal{X}''_i$.
    2.) Generate adversarial data $\mathcal{X}_{adv,i}$ from $\mathcal{X}''_i$.
    3.) Train local model $\mathcal{M}_{\theta_{i, t}}$ on the pooled $\mathcal{X}'_{i} \cup \mathcal{X}_{adv,i}$
    4.) Send trained model's weights $\theta_{i,t}$ to server.
    5.) Aggregate all the clients' weights, resulting in $\theta_{i,t+1}$.
    6.) Send back $\theta_{i,t+1}$ to the clients for the next round $t+1$.
    }
    \label{fig:workflow}
\end{figure*}

%
%

To tackle the challenges discussed in Section \ref{sec:introduction}, we introduce a novel federated learning pipeline for SER that is robust towards white-box attacks on speech data (\cf \autoref{fig:workflow}).
Under the assumption that each client already has its own speech data, our pipeline consists of three components. First, we employ a data pre-processing module that extracts the log Mel spectrograms from speech signals of each client. The log Mel spectrograms are further used as the input of SER models. Second, we propose a federated adversarial learning module, which is the defence strategy at the training time. Specifically, in the federated adversarial training module, the SER models learnt by FL in Section~\ref{sec:fl} are defended with adversarial training illustrated in Section~\ref{sec:adv_training}. Finally, a randomisation module for defence at the inference time is proposed to further improve the SER models' robustness with the approach of randomisation outlined in Section~\ref{sec:randomisation}.
In the following, we will give details about each of the three modules. 

\subsection{Data Pre-processing} 
As stated previously, our pipelines assumes that each participating client $\mathcal{F}_{i}$ to the FL has its own separate audio data $\mathcal{D}_{i}$.
A pre-processing step is necessary before feeding the data to the local model. Log Mel spectrograms are extracted from each client's audio data. Compared to regular spectrograms, log Mel spectrograms use the Mel scale on the $y$ axis instead of 
linearly scaled 
frequency.As a result, each client $\mathcal{F}_{i}$ would have its own Log Mel spectrogram data $\mathcal{X}_{i}$ that will be used as input to its local model $\mathcal{M}_{\theta_{i, t}}$.





\subsection{Defence at Training Time -- Adversarial FL} 
In \autoref{fig:workflow} $(a)$, adversarial federated learning is the first stage of defence within our pipeline. To avoid training more complex neural networks and the obfuscated gradient effects~\cite{DBLP:conf/icml/AthalyeC018}, we do not use 
a denoising technique. The adversarial training applied in this work is vanilla adversarial training, which provides better performance on original data and comparable performance on adversarial data compared to similarity-based adversarial training~\cite{9054087}.

For our adversarial learning module, we generate adversarial samples using white-box attacks.
To obtain adversarial samples, the input is $\mathcal{X}_i$ and $\mathcal{Y}_i$, the original samples and labels from client $\mathcal{F}_i$ (for a client $i\in[1,N]$). 
Adversarial federated learning follows the same training principle outlined for vanilla federated learning. 
However, at the beginning of each round $t$,
training data samples are divided randomly to $\mathcal{X}'_i$ and $\mathcal{X}''_i$ with equal number of samples, this ensures having different set of samples in each round. 

Using the local model of the last round $\mathcal{M}_{\theta_{i,t}}$, white-box attacks are applied to $\mathcal{X}''_i$ , generating the adversarial samples $\mathcal{X}_{adv,i}$. 
Subsequently, the union of $\mathcal{X}'_i \cup \mathcal{X}_{adv,i}$ is used for training.
The loss function is defined as follows:
\begin{multline} \label{eq:vanilla_advTrain}
\mathcal{\hat L}(\theta_i, \mathcal{X}'_i \cup \mathcal{X}_{adv,i}, \mathcal{Y}_i) = \alpha \cdot \mathcal{L}(\theta_i, \mathcal{X'}_i, \mathcal{Y}_i) + \\ (1 - \alpha) \cdot \mathcal{L}(\theta, \mathcal{X}_{adv,i}, \mathcal{Y}_i),
\end{multline}
where $\alpha$ is a constant value used to adjust the ratio of loss value based on the original data samples $\mathcal{L}(\theta_i, \mathcal{X}'_i, \mathcal{Y}_i)$ and adversarial samples $\mathcal{L}(\theta_i, \mathcal{X}_{adv,i}, \mathcal{Y}_i)$. %
Then, when the local training is complete, the new weights $\theta_{i,t}$ are sent to the FL server for the next round $t+1$.
Given that we need a trained model to generate adversarial samples, in the first round, we only use the original samples $\mathcal{X}_i$ for each respective client $i \in [1, N]$, and begin our adversarial training from the second round onwards.

%

\subsection{Defence at Inference Time -- Randomisation} \label{sec:defence_at_inference_time}
As depicted in \autoref{fig:workflow} $(b)$, two randomisation layers at the inference time are applied: random resizing and random padding. The random resizing layer randomly enlarges the input log Mel spectrogram with the dimension of $(W, H, 1)$ to the dimension of $(W', H', 1)$, where $W'$ and $H'$ are randomly chosen from a reasonably small range, which means $|W' - W|$ and $|H' - H|$ are small: 
\begin{equation} \label{eq:randomisation}
(W, H, 1) \xrightarrow[\mbox{resizing}]{\mbox{random}} (W', H', 1) \xrightarrow[\mbox{padding}]{\mbox{random}} (W'', H'', 1).
\end{equation}
By fine-tuning the hyperparameters for randomisation, we pursue two goals. 
First, randomisation has a minimal impact on the  models' performance on the original data.
Second, it mitigates the adversarial attacks' effects on the models' performance. 
Using a small enough range has been shown to have no significant effect on model performance on the original dataset~\cite{xie2018mitigating}.
Next, a random padding layer fills in certain values (\eg zero) around the resized spectrograms, resulting in the final dimension $(W'', H'', 1)$. The two computationally efficient randomisation layers make the models more robust against adversarial samples generated using one-step and iterative attacks.

\begin{figure}[h]
    \centering
    \includegraphics[trim={0 0 .8cm 0},clip,width=.9\columnwidth]{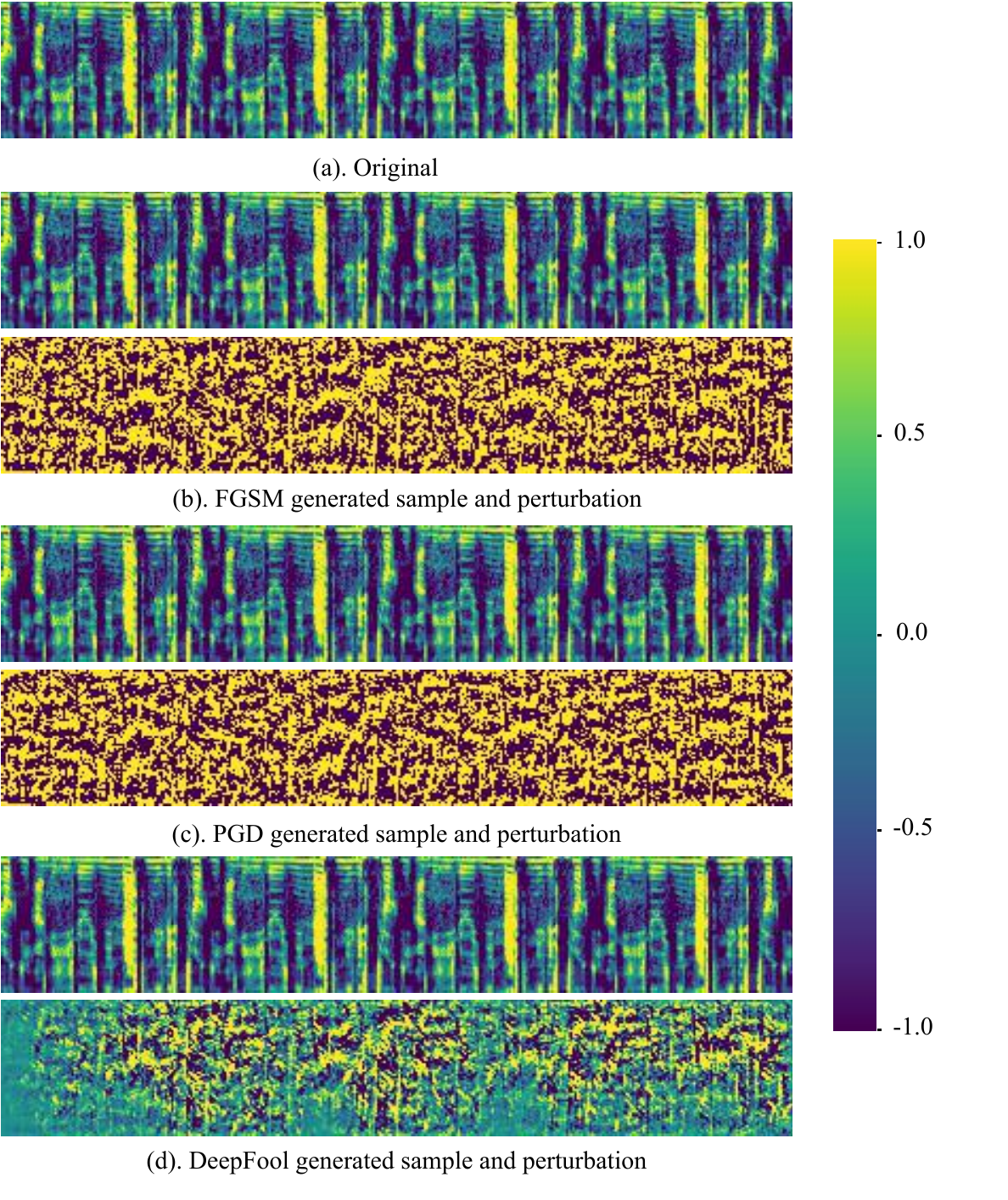}
    \caption{One example of a log Mel spectrogram from the DEMoS training dataset to visualise the adversarial attacks. For each log Mel spectrogram, the time length is 373 frames and the max Mel frequency is 64. The model applied herein is one converged general federated learnt model after 300 rounds. 
    The values of the perturbations are multiplied by 100 for easy observation.}
    \label{fig:deepfool}
\end{figure}





\section{Experiments and Results} \label{sec:experiments_and_results}

In this section we evaluate our pipeline using the Database of Elicited Mood in Speech (DEMoS)~\cite{parada2019demos}, a popular SER dataset.
First we shall discuss the properties of this dataset and our approach for obtaining a data split suitable for FL. We give more details about this dataset in \autoref{subsec:dataset}.
Then, in \autoref{subsec:exp_setup}, we describe our experimental setup, including details on how we extracted the log Mel spectrograms from the audi data, model architectures, the setting used for adversarial federated training and randomisation, reproducibility, and evaluation metrics.  
Finally we discuss our experimental results in \autoref{subsec:exp_results}. 

\subsection{Dataset} \label{subsec:dataset}

The DEMoS~\cite{parada2019demos} is an Italian emotional speech corpus of $7.7$ hours of audio recordings, 
collected from $68$ speakers (23 females and 45 males). 
The speakers' emotion were induced by an arousal-valence progression. In total, $9,365$ emotional and $332$ neutral speech samples were recorded. As the neutral state is a minority class, it is not considered in our experiment. 
Therefore, we employ the $9,365$ emotional speech samples (average duration: $2.86$\,seconds $\pm$ standard deviation: $1.26$\,seconds) annotated into seven classes, including \emph{anger}, \emph{disgust}, \emph{fear}, \emph{guilt}, \emph{happiness}, \emph{sadness}, and \emph{surprise}. 

\begin{table}[ht]
\centering
\caption{Emotion distribution of the applied DEMoS dataset.
}
\begin{tabular}{P{1cm}|P{1cm}|P{1cm}|P{1cm}}
\toprule
\# & Train & Test & $\sum$ \\
\hline
Anger &1,155 &\ \ 322 &1,477 \\
\hline
Disgust &1,354 &\ \ 324 &1,678 \\
\hline
Fear &\ \ 927 &\ \ 229 &1,156 \\
\hline
Guilt &\ \ 898 &\ \ 231 &1,129 \\
\hline
Happiness &1,127 &\ \ 268 &1,395 \\
\hline
Sadness &1,228 &\ \ 302 &1,530 \\
\hline
Surprise &\ \ 802 &\ \ 198 &1,000 \\
\hline
$\sum$ &7,491 &1,874 &9,365 \\
\bottomrule
\end{tabular}
\label{tab:demos}
\end{table}

In our experiment, all audio recordings were sampled with $16$\,kHz. 
Because federated learning aims to perform on each speaker's data in a long-term manner for personalised SER-related applications, we divide the database into a training set and a test set with a speaker-dependent strategy rather than the `classic' speaker-independent one. Each actor's speech is divided into an 80\,\% training set and a 20\,\% test set, each of which composes the global training set and test set, respectively. The emotion distribution is described in Table~\ref{tab:demos}.

\subsection{Experimental Setup}\label{subsec:exp_setup}

\textbf{Log Mel Spectrograms Extraction:}
We set the audio sample length as the maximum one as 5.884\,seconds, which means shorter audio samples will be self-repeated. 
We set the sliding window as 512 time frames, the overlap as 256 time frames, and the Mel bins as 64
in number 
for the log Mel spectrogram extraction~\cite{ren2020generating}. 
As a result, the generated log Mel spectrograms share the dimensions of (373, 64), where 373 is the dimension along the time steps, and 64 is the 
number of 
Mel frequencies. 

\begin{figure}[!t]
    \centering
    \includegraphics[width=.20\textwidth]{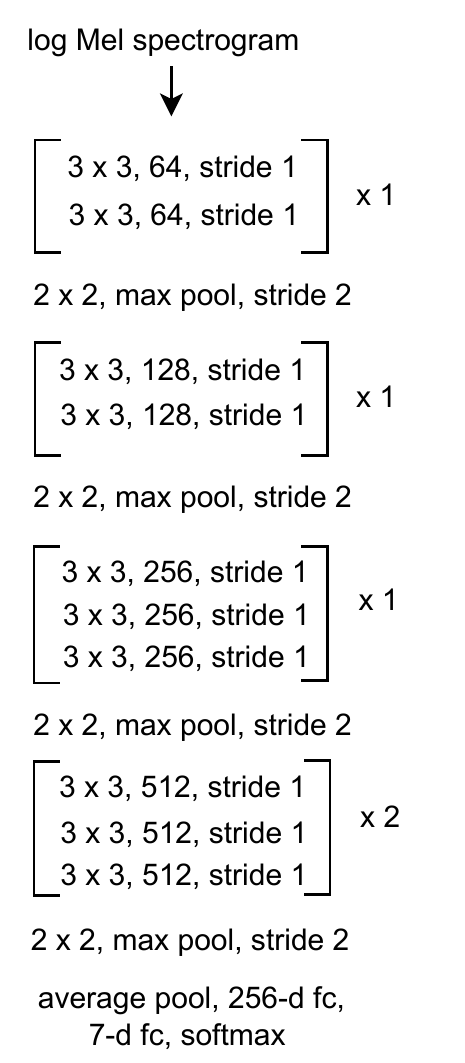}
    \caption{The VGG-15 architecture. Specifically, building blocks are in square brackets with numbers of stacked blocks and every convolutional layer in the square brackets is followed by a batch normalisation layer and a `ReLU' activation function.}
    \label{fig:models}
\end{figure}

\textbf{Model's Architecture:}
VGG and VGG-like architectures have been successfully applied in classifying spectrogram images for a number of audio tasks
~\cite{10.3389/fdgth.2021.799067, 9054087}. Therefore, in this work, we trained VGG-15 models in the federated setting.
As shown in \autoref{fig:models}, the architecture of VGG-15 consists of five convolutional blocks with the output channel numbers of 64, 128, 256, 512 and 512, 
each of which is followed by a local max pooling layer with a kernel size of $(2, 2)$. Each convolutional layer is followed by a batch normalisation layer and a `ReLU' activation function~\cite{relu} to stabilise and accelerate the training process~\cite{Ren2018AttentionbasedCN}. Before the final two fully connected layers for the final classification, a global average pooling layer is applied.

\textbf{Adversarial Federated Training:}
In order to perform adversarial training, adversarial data should be generated first using one of the white-box attacks. For the iterative attack DeepFool, the maximum iteration for the optimisation procedure in is set to 5, since the study~\cite{moosavi2016deepfool} found that DeepFool empirically converges in less than 3 iterations for perturbation to fool the classifier. For better comparison, we also set the maximum iteration for PGD to 5. Moreover, we set the norm of the generated perturbation for FGSM, PGD and DeepFool as $l_{\infty}$, $l_{\infty}$ and $l_{2}$ respectively. The $\epsilon$ in Equation~(\ref{eq:FGSM}), $\eta$ in Equation~(\ref{eq:PGD}) and $\zeta$ in Equation~(\ref{eq:DeepFool}) are set to 0.05, 0.05 and 0.02 respectively. In addition, the settings for the adversarial attack are the same for training and testing. Some generated adversarial examples are visualised in \autoref{fig:deepfool}. It can be seen that DeepFool tends to generate less perceptible attacks to fool the classifier compared to FGSM and PGD.

We set the $\alpha$ in Equation~(\ref{eq:vanilla_advTrain}) to 0.5, which means half of the original samples $\mathcal{X}_{i}$ from the client $\mathcal{F}_{i}$ (where $i\in[1,N]$ represents the client) are randomly selected from $\mathcal{D}_{i}$ to generate the corresponding adversarial examples $\mathcal{X}_{adv,i}$ along with the remaining half of the original samples to train the model to have a fair comparison with the vanilla federated learnt models.

The batch size for training is set to 8 to account for limited memory, while for validation and testing it is set to 1 to maximise the ability to randomise. The optimiser \say{Adam} \cite{kingma2014adam} is applied with a fixed learning rate of 0.001.


For the implementation of FL we use our framework based on the Flower FL framework \cite{beutel2020flower} (an open source framework for Federated Learning), where we have adapted Docker to containerise our FL clients and FL servers. This means that each FL client runs in a Docker container and communicates with the FL server container. Once the FL training is complete, a model with the aggregated weights of the last round is stored for evaluation.

\textbf{Randomisation:}
The original dimension of the generated log Mel spectrogram is (373, 64, 1). In Equation~(\ref{eq:randomisation}), the $W'$ is randomly chosen in [373, 380), and $H'$ is randomly chosen in [64, 66). After the resizing, we apply the random padding on the boundaries of the log Mel spectrogram by the value 0.5 and after the padding, the final $(W'', H'', 1)$ dimension is (380, 66, 1). We fine-tune the hyper-parameters (\ie the range that $W'$ are $H'$ chosen from, and values of $W''$ and $H''$) of the randomisation based on the set-aside validation dataset. 

\textbf{Reproducibility:}
The open-source Adversarial Robustness Toolbox library (ART) \cite{art2018} is used to generate the adversarial data.
In terms of hardware, the experiments are conducted on an Nvidia DGX machine\footnote{https://www.nvidia.com/en-us/data-center/dgx-station-a100}, which has 8 $\times$ NVIDIA A100 GPUs with a total of 320\,GB GPU memory and an AMD 7742 64 cores 2.25\,GHz CPU with 512\,GB of RAM. DGX is a powerful machine that allowed us to run the FL experiments.

\textbf{Evaluation Metrics:}
Compared with accuracy (\ie weighted average recall), Unweighted Average Recall (UAR) can better evaluate models' classification performance on imbalanced datasets~\cite{uar1}. UAR is used to further evaluate the converged federated models' performance on data where randomisation operations have been applied, adversarial examples  and randomised adversarial examples. 





\begin{figure}[!t]
    \centering
    \includegraphics[width=.45\textwidth]{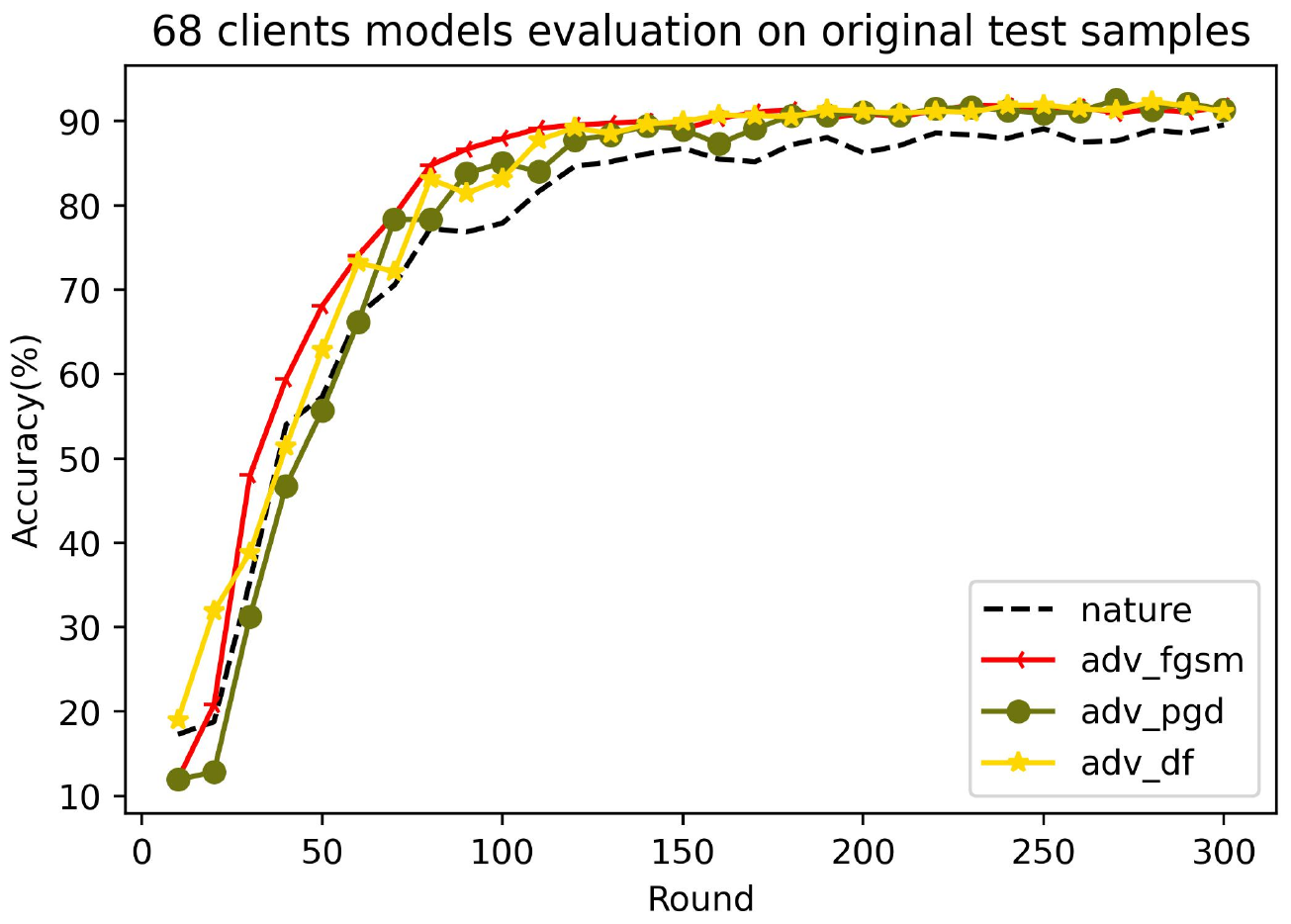}
    \caption{Performance comparison on test data between a vanilla federated learnt model, and federated adversarial trained models with the DeepFool, FGSM and PGD algorithms. The models are evaluated at each $10^{th}$ round. Specifically, in the legends, `Nature' represents the vanilla federated learnt model; `AdvTrain' means the adversarial training; `Attk.' means adversarial attack.}
    \label{fig:68_models}
\end{figure}

\subsection{Experimental Results} \label{subsec:exp_results}

\autoref{fig:68_models} illustrates the UAR score of the trained FL models, comparing the nature federated learnt model and the federated adversarial trained models with PGD, FGSM  and DeepFool, respectively. 
The models are evaluated on the original (unperturbed) test data every 10 rounds.
The four mentioned models converge to a UAR score of around 90\,\% with diverse convergence times. We can see that, the three federated adversarial training algorithms share similar convergence speed with the nature model, where all models herein converge after 150 rounds. Therefore, we conduct further tests for the above models after 150 rounds until the final 300 rounds, every 10 rounds.

\begin{figure*}
     \centering
     \begin{subfigure}[b]{0.46\textwidth}
         \centering
         \includegraphics[width=\textwidth]{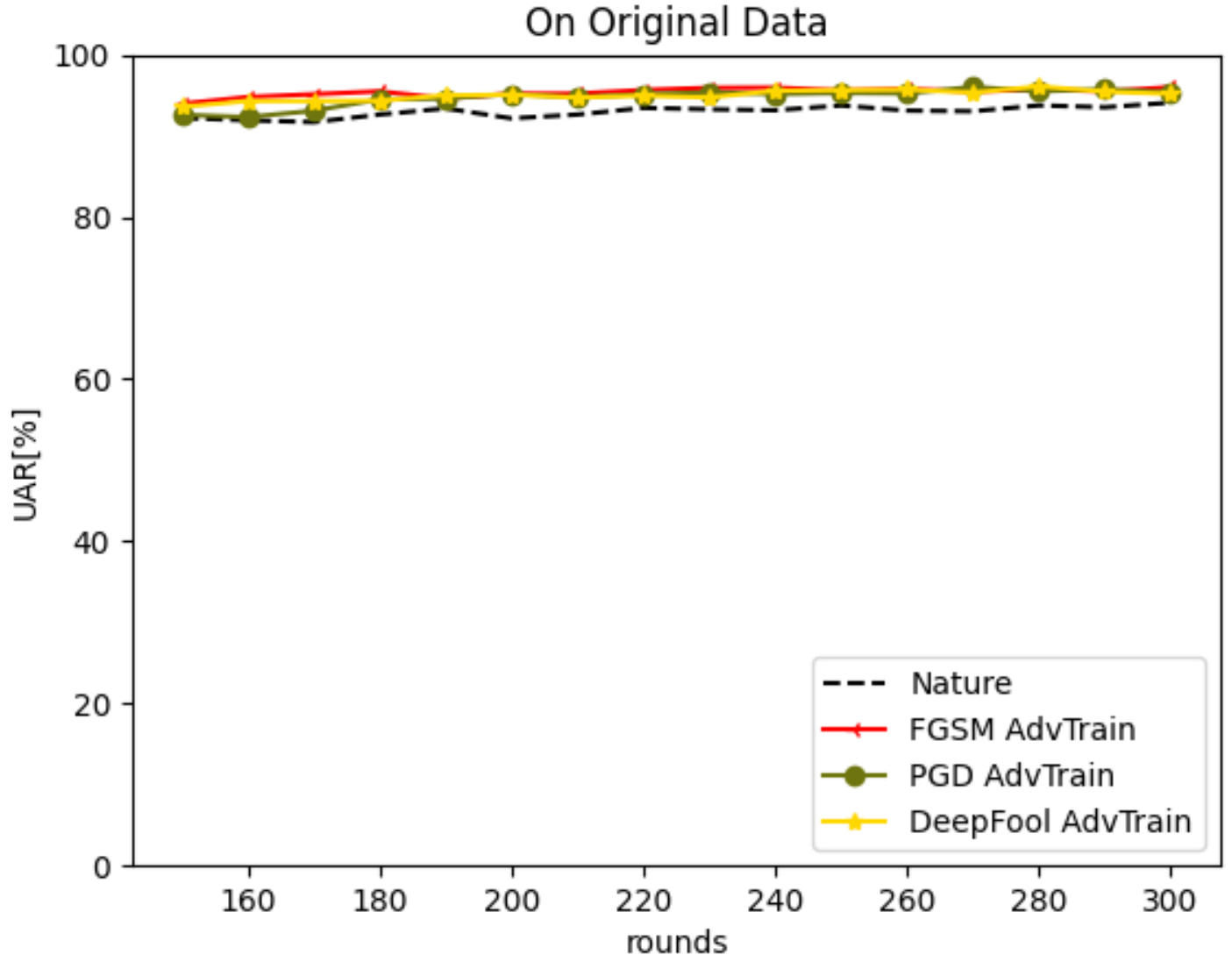}
         \caption{}
         \label{fig:demos1}
     \end{subfigure}
     \hfill
     \begin{subfigure}[b]{0.46\textwidth}
         \centering
         \includegraphics[width=\textwidth]{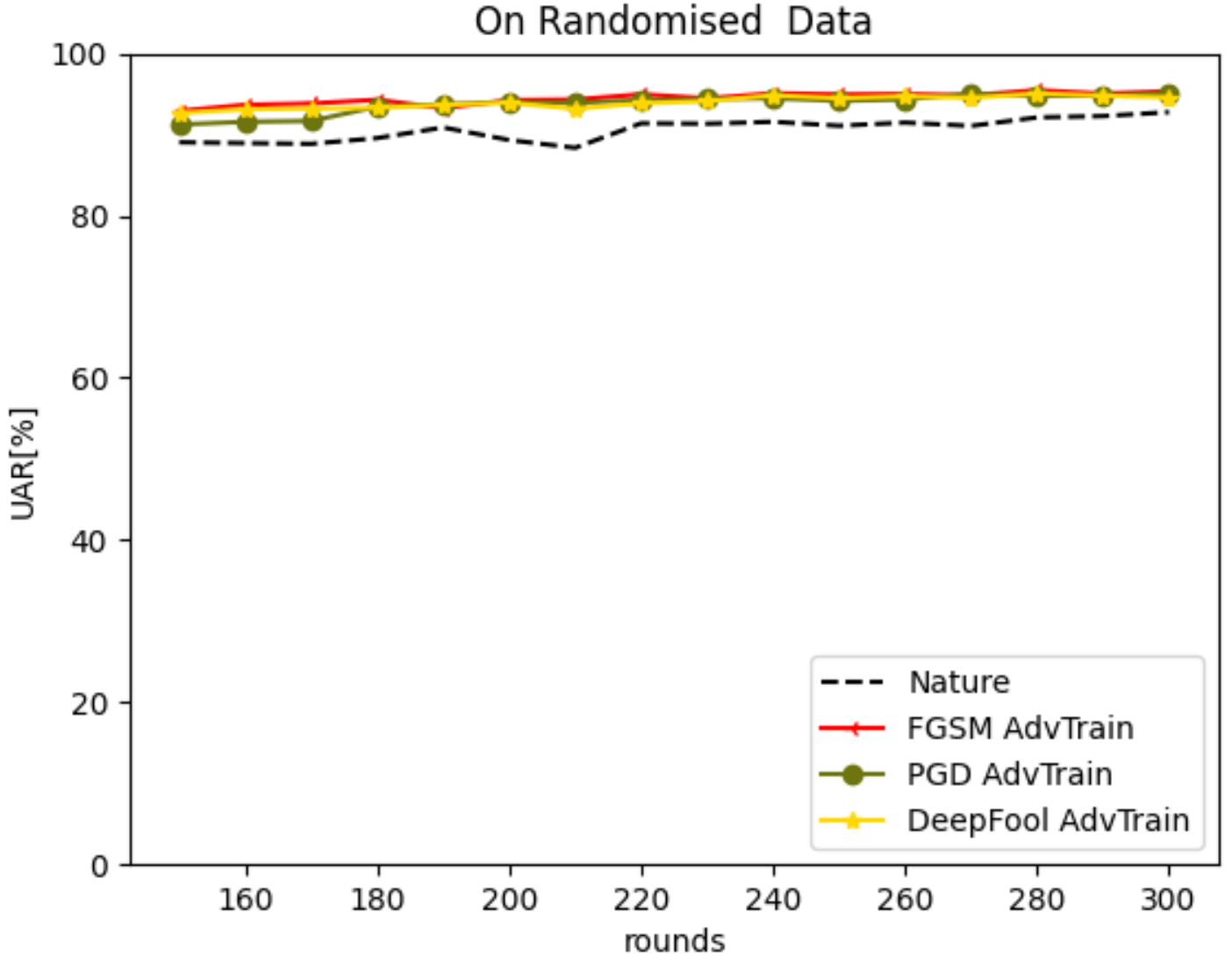}
         \caption{}
         \label{fig:demos2}
     \end{subfigure}
     \hfill
     \begin{subfigure}[b]{0.46\textwidth}
         \centering
         \includegraphics[width=\textwidth]{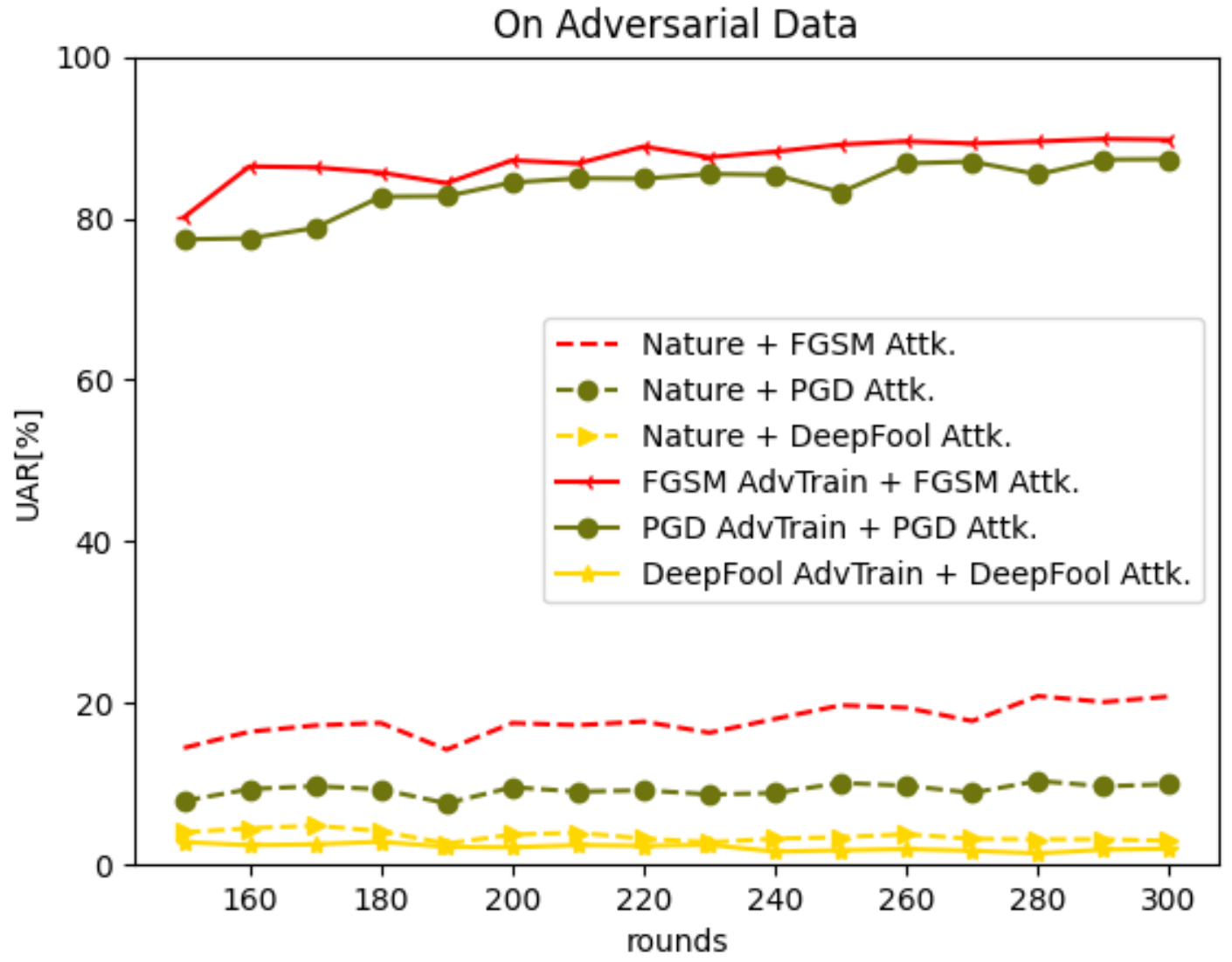}
         \caption{}
         \label{fig:demos3}
     \end{subfigure}
     \hfill
     \begin{subfigure}[b]{0.46\textwidth}
         \centering
         \includegraphics[width=\textwidth]{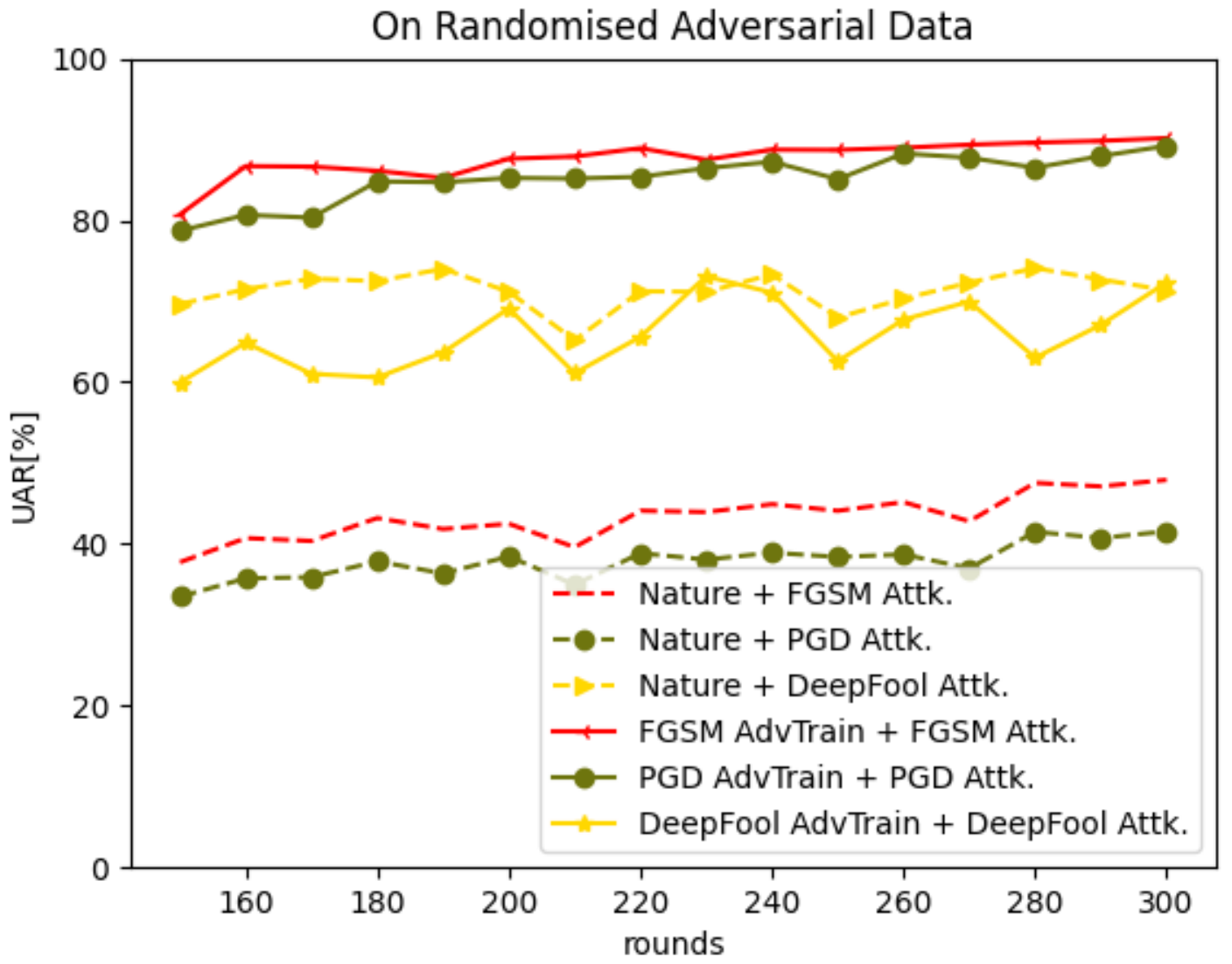}
         \caption{}
         \label{fig:demos4}
     \end{subfigure}
        \caption{Performance on DEMoS every 10 rounds until 300 rounds. (a) The results of models tested on the original test dataset; (b) The results on the randomised test dataset; (c) The results on the adversarial samples; (d) The performances on the randomised adversarial samples. Specifically, in the legends, `Nature' represents the federated learnt model; `AdvTrain' means the adversarial training; `Attk.' means adversarial attack.}
        \label{fig:demos_res_tab}
\end{figure*}

From \autoref{fig:demos_res_tab}, there are three major findings. Firstly, \autoref{fig:demos1} and \autoref{fig:demos2} show that adversarial learnt models perform better than nature models; the randomisation layer herein itself does not significantly reduce the model's performance. Secondly, from \autoref{fig:demos3}, we can see that our adversarial training strategy is successful, helping models fight against the adversarial attacks, besides DeepFool. 
%
%
Thirdly, by comparing \autoref{fig:demos3} and \autoref{fig:demos4}, we can observe that randomisation can further help mitigate the effects of adversarial attack, especially for nature models under FGSM, PGD and DeepFool attacks and adversarial federated learnt models under DeepFool attacks. 
However, we do not see an improvement for the federated adversarial models under FGSM and PGD attacks after randomisation.
%
Our evidence supports that
adversarial training is the key step in defending the models against the FGSM and PDG attacks under our settings, which have strong transferability. %
On the other hand,  DeepFool is more likely to over-fit on the target models, and thus randomisation can better help to destroy the structure of the perturbations.

\begin{figure}
     \centering
     \begin{subfigure}[b]{0.45\textwidth}
         \centering
         \includegraphics[width=\textwidth]{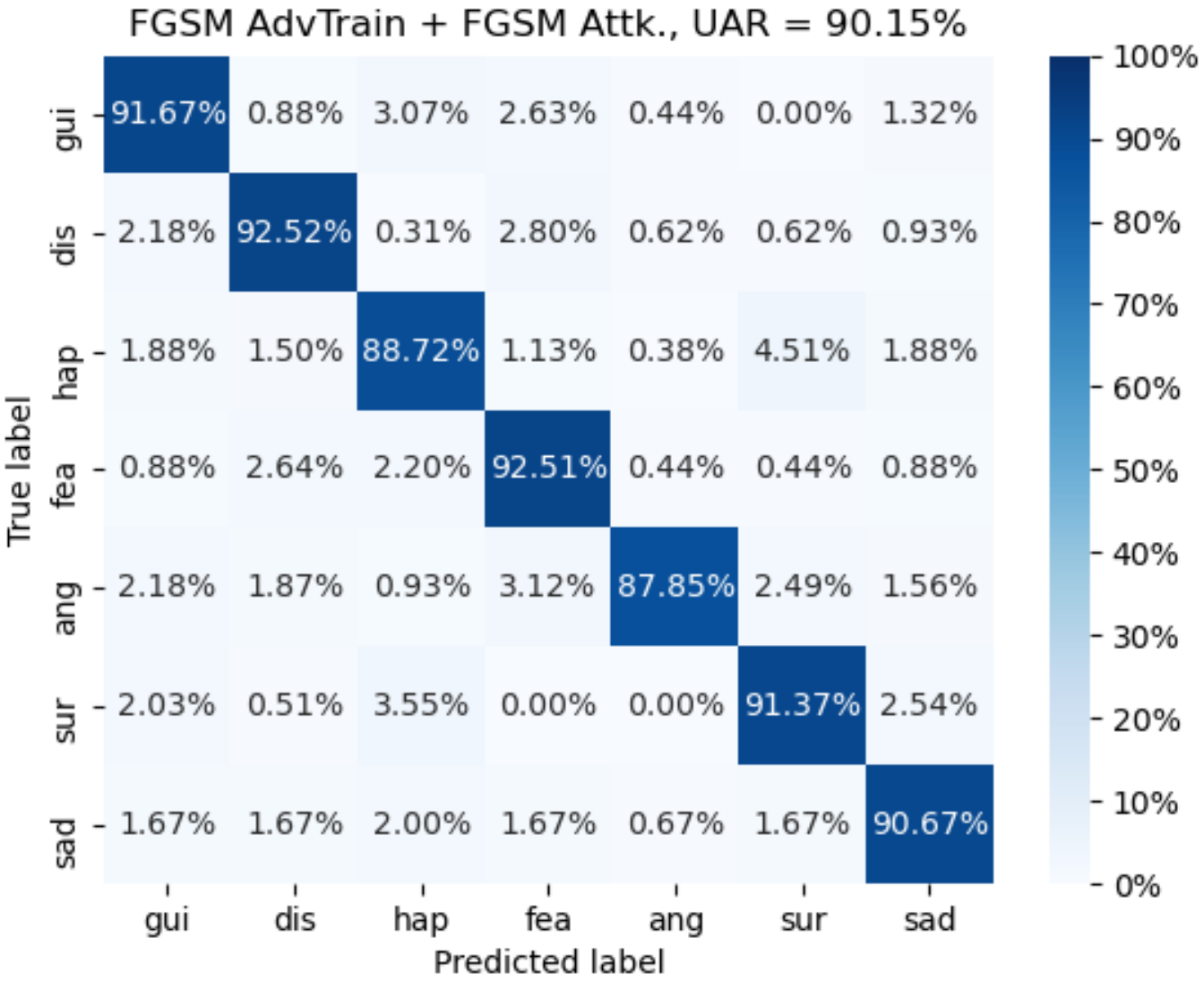}
         \caption{}
         \label{fig:cf_fgsm}
     \end{subfigure}
     \hfill
     \begin{subfigure}[b]{0.45\textwidth}
         \centering
         \includegraphics[width=\textwidth]{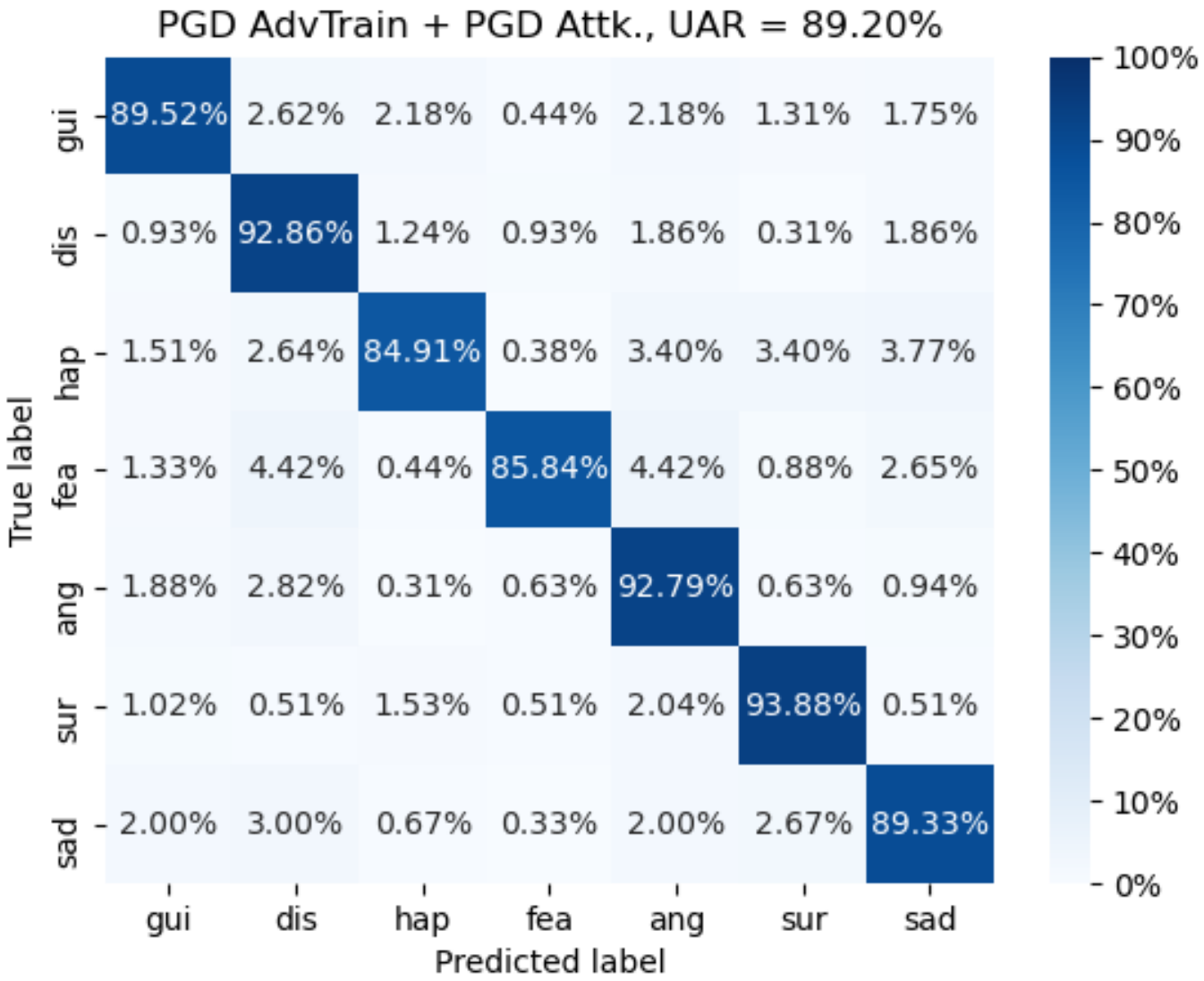}
         \caption{}
         \label{fig:cf_pgd}
     \end{subfigure}
     \hfill
     \begin{subfigure}[b]{0.45\textwidth}
         \centering
         \includegraphics[width=\textwidth]{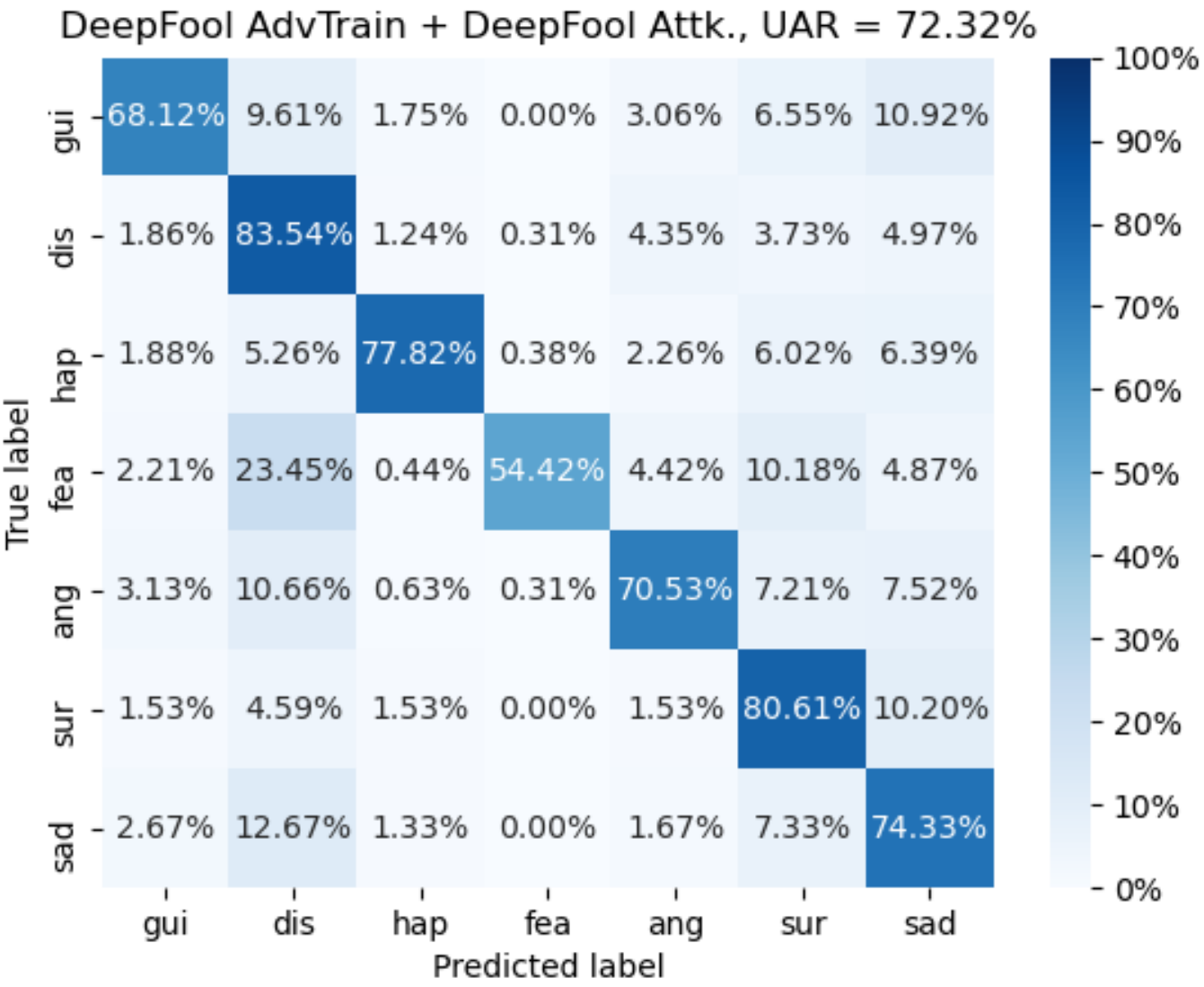}
         \caption{}
         \label{fig:cf_deepfool}
     \end{subfigure}
        \caption{Confusion matrices for adversarial training models under adversarial attacks after $300$ rounds on the DEMoS test dataset. `gui': Guilt, `dis': Disgust, `hap': Happiness, `fea': Fear, `ang': Anger, `sur': Surprise, `sad': Sadness.}
        \label{fig:cf_demos}
\end{figure}

\begin{table}[h]
\caption{The performance (UAR [\%]) of nature and adversarial federated learnt models after 300 rounds on the test dataset, when it is deceived by adversarial attacks. The type of adversarial attack is consistent with the one for generating the adversarial examples for adversarial training.}
\scalebox{0.9}{\begin{tabular}{c|c|rrrr}
\toprule
Models & Test data & \multicolumn{4}{c}{Attacks} \\ \hline
\multicolumn{1}{l|}{} & \multicolumn{1}{l|}{} & \multicolumn{1}{c|}{No} & \multicolumn{1}{c|}{FGSM} & \multicolumn{1}{c|}{PGD} & \multicolumn{1}{c}{DeepFool} \\ \hline
\multirow{4}{*}{Nature} & Original & \multicolumn{1}{r|}{94.08} & \multicolumn{1}{r|}{-} & \multicolumn{1}{r|}{-} & - \\ \cline{2-6} 
 & Randomised & \multicolumn{1}{r|}{92.75} & \multicolumn{1}{r|}{-} & \multicolumn{1}{r|}{-} & - \\ \cline{2-6} 
 & Adversarial & \multicolumn{1}{r|}{-} & \multicolumn{1}{r|}{20.82} & \multicolumn{1}{r|}{9.96} & 2.96 \\ \cline{2-6} 
 & Randomised adversarial & \multicolumn{1}{r|}{-} & \multicolumn{1}{r|}{47.91} & \multicolumn{1}{r|}{41.55} & 71.34 \\ \hline
\multirow{4}{*}{\begin{tabular}[c]{@{}c@{}}Adversarial\\ federated \\ learnt\end{tabular}} & Original & \multicolumn{1}{r|}{-} & \multicolumn{1}{r|}{95.99} & \multicolumn{1}{r|}{95.40} & 95.16 \\ \cline{2-6} 
 & Randomised & \multicolumn{1}{r|}{-} & \multicolumn{1}{r|}{95.32} & \multicolumn{1}{r|}{95.10} & 94.56 \\ \cline{2-6} 
 & Adversarial & \multicolumn{1}{r|}{-} & \multicolumn{1}{r|}{89.72} & \multicolumn{1}{r|}{87.31} & 1.98 \\ \cline{2-6} 
 & Randomised adversarial & \multicolumn{1}{r|}{-} & \multicolumn{1}{r|}{90.15} & \multicolumn{1}{r|}{89.20} & 72.32 \\ \bottomrule
\end{tabular}}
\label{tab:result_fl}
\end{table}

From Table~\ref{tab:result_fl}, there are several observations. Firstly, for both the nature models and federated adversarial trained models, randomisation is successful in mitigating the adversarial attacks. When we apply randomisation onto the original test dataset, the performance only drops by a small amount (usually less than absolute 1\,\% on UAR). %
When randomisation is applied on the generated adversarial examples, it improves considerably the performance. 
This is because randomisation probably destroys the specific adversarial perturbations pattern~\cite{xie2018mitigating}. 
Specifically, randomisation helps the models mitigate the DeepFool attack, improving the UAR from 
2.96\,\% to 71.34\,\% ($p < 0.001$ in a one-tailed z-test) for the nature model and from 1.98\,\% to 72.32\,\% for the adversarial federated learnt  model ($p < 0.001$ in a one-tailed z-test). Moreover, the UAR improvements for nature models under FGSM (from 20.82\,\% to 47.91\,\%) and PGD (9.96\,\% to 41.55\,\%) attacks are also significant ($p < 0.001$ in a one-tailed z-test). 

Secondly, with adversarial training, models' performances under the FGSM (20.82\,\% to 89.72\,\%) and PGD (9.96\,\% to 87.31\,\%) attacks improve a lot ($p < 0.001$ in a one-tailed z-test).
This result is consistent with existing work in the literature, where single-step attacks (\eg FGSM) and PGD were shown to be more transferable and less prone towards over-fitting~\cite{xie2018mitigating}.
%
%
However,  we observe that the adversarial training does not help the models to mitigate the DeepFool adversarial effects. 
The reason is that DeepFool iterates until the generated attack samples successfully fool the classifier, in which process DeepFool is more likely to over-fit on the specific models, and has less transferability~\cite{xie2018mitigating}. 

The confusion matrices of adversarial federated learnt models' performances on the randomised adversarial test dataset after 300 rounds can be referred to in \autoref{fig:cf_demos}. From \autoref{fig:cf_fgsm} and \autoref{fig:cf_pgd}, we can see that most test samples are labeled correctly. 
However, in \autoref{fig:cf_deepfool},  many samples are misclassified as the Disgust emotion -- especially the \emph{fear} emotion. This can be caused by the somewhat similar acoustic characteristics of these negative feelings.

\section{Discussion and Future Work} \label{sec:discussion}

Our study demonstrates a novel pipeline that can be used to efficiently mitigate adversarial white-box attack on SER data. In this section we consider: the main takeaways from our work with respect to the performance of our two layer defence against adversarial attacks; a comparison with work on centralised training on DEMoS; and finally, considerations regarding future work that could be conducted with our pipeline.

\textbf{White-box attack comparison:}
In our work, we analyse two types of adversarial white-box attacks, single-step attacks (\ie FGSM) and iterative attacks (\eg DeepFool). Overall, our findings are consistent with previous work on adversarial attacks. 

After conducting the experiments, we come to the conclusion that single-step attacks are weaker than iterative attacks, but more transferable~\cite{xie2018mitigating}.
By comparing the effects of adversarial training on defending against single-step and iterative attacks,
we observe that adversarial training mitigates better the effects of single-step attacks (\ie FGSM) than iterative attacks (\eg DeepFool). We believe that the cause for this is due to the nature of the iterative attacks, which perform multiple iterations to find the minimal perturbations that can fool the classifier. 
In other words, perturbations generated by iterative attacks tend to over-fit on the parameters of the specific classifier ~\cite{xie2018mitigating}, which makes the iterative attacks more powerful. Thus, adversarial training doesn't effectively defend against the iterative attacks. 

As a result, we propose the randomisation -- the second defence strategy within our pipeline --  to be more effective, via utilising resizing and padding by random proportions on the test data at inference time. By this, randomisation tends to destroy the specific structures of perturbations. 
Our experimental results agree on the previous statement, and show that the added randomisation makes the models more robust against iterative attacks, where DeepFool attacks are applied.

\textbf{Comparison with centralised training:}
The VGG model architecture that we use in our experiments is similar to the one utilised in the study of~\cite{ren2020generating}. The use-case was for the centralised adversarial training on the DEMoS corpus in order to reduce the impact of the FGSM attack where a highest UAR -- 86.7\,\% was achieved on the unperturbed original test data and 82.1\,\% on the perturbed test data. 
However, in that work, data is split to training, developing \&testing, in a speaker-independent way, where training data contains a set of speakers that were not present in the test data. 

In our work, on the other hand, the data is split in a speaker-dependent way. 
As a result of the splitting strategy that our FL workflow restricts by design, a direct comparison between our results and those from~\cite{ren2020generating} cannot be made. 

\textbf{Future work:}
In the future, our work can be extended and further evaluated, below we discuss three directions that we are particularly keen to explore.
%
Firstly, in real-life communication scenarios, clients (\eg mobile devices) are distributed in various environments and with different languages. And potentially, each client can have a different dataset. This raises the research question of how our framework will perform and how robust can the learnt FL model be in such heterogeneous environment.

Secondly, if attackers use an ensemble of adversarial examples generated by different algorithms, how does our workflow defend against the unseen adversarial attacks? and to what extend does it generalise over these new attacks. 

Thirdly, there are also numerous opportunities for applying our pipeline within a different domain. For instance, in the context of intelligent industrial production, where the datasets are recorded on industrial production machines (\eg computer numerical control machines) via acoustic and vibration analytic devices. How transferable is our federated adversarial learning framework to this domain, and how does it perform and defend against adversarial attacks in such a scenario?






\section{Conclusion} \label{sec:conclusion}
In this article, we proposed a framework that is composed of two defence stages, the first is the federated adversarial training that  at the training time and the other one at the inference time, to mitigate white-box adversarial effects on the federated learnt speech emotion recognition models. We conducted experiments on the database of elicited mood in speech using the VGG-15 architecture against three attack methods. The experimental results indicated that adversarial federated learning can better fight against attacks with higher transferability and that randomisation works better on stronger iterative DeepFool attacks. By combining the two defence strategies together, we achieved significant improvement under adversarial attacks. In particular, the unweighted average recalls on the randomised adversarial test dataset were 90.15\,\% for the fast gradient sign method attack, 89.20\,\% for the projected gradient attack and 72.32\,\% for the DeepFool attack. 
These findings make federated learning in speech emotion recognition appear a sound option to help collect sufficient experience for tomorrow's engines in real-world applications.

\section*{Acknowledgement}
This work was partially supported by the research projects ``IIP-Ecosphere'', granted by the German Federal Ministry for
Economics and Climate Action (BMWK) via funding code No.\,01MK20006A, 
%
and ``LeibnizKILabor'', granted by the German Federal Ministry of Education and Research (BMBF) via funding code No.\,01DD20003.

\bibliographystyle{IEEEtran}
\bibliography{bibliography}  




\end{document}